# Electrical valley filtering in transition metal dichalcogenides

Tzu-Chi Hsieh[1], Mei-Yin Chou[1,2,3], and Yu-Shu Wu[4,5]

[1] Institute of Atomic and Molecular Sciences, Academia Sinica, Taipei 10617, Taiwan
[2] School of Physics, Georgia Institute of Technology, Atlanta, Georgia 30332, USA
[3] Department of Physics, National Taiwan University, Taipei 10617, Taiwan
[4] Department of Electrical Engineering, National Tsing-Hua University, Hsinchu 30013, Taiwan
[5] Department of Physics, National Tsing-Hua University, Hsinchu 30013, Taiwan

This work investigates the feasibility of electrical valley filtering for holes in transition metal dichalcogenides. We look specifically into the scheme that utilizes a potential barrier to produce valley-dependent tunneling rates, and perform the study with both a $k \cdot p$ based analytic method and a recursive Green's function based numerical method. The study yields the transmission coefficient as a function of incident energy and transverse wave vector, for holes going through lateral quantum barriers oriented in either armchair or zigzag directions, in both homogeneous and heterogeneous systems. The main findings are the following: 1) the tunneling current valley polarization increases with increasing barrier width or height, 2) both the valley-orbit interaction and band structure warping contribute to valley-dependent tunneling, with the former contribution being manifest in structures with asymmetric potential barriers, and the latter being orientation-dependent and reaching maximum for transmission in the armchair direction, and 3) for transmission ~ 0.1, a tunneling current valley polarization of the order of 10% can be achieved.

## I. INTRODUCTION

Valleytronics has recently attracted a lot of attention [1, 2]. It can be realized in 2D hexagonal structures such as graphene [3] or monolayer transition metal dichalcogenides (TMDCs) [4] where electrons carry valley pseudospin – a binary quantum index in association with the existence of two inequivalent and degenerate band structure valleys at the corners (K and K′) of Brillouin zone. The large crystal momentum separation between K and K′ protects the valley pseudospin well from inter-valley scattering and leads to good-sized valley coherence suitable for valley-based information processing.

The implementation of valleytronics requires valley control. Various schemes to manipulate valley polarization have been proposed, such as those based on ballistic transport through a zigzag nanoribbon [1], defect scattering [5], band structure warping [6], and strain [7]. In gapped graphene or TMDCs, where the presence of inversion symmetry breaking leads to the emergence of opposite orbital magnetic moments for electrons in the two valleys [2], alternative valley control are made available by coupling the valley pseudospin to external fields, e.g., out-of-plane magnetic fields or in-plane electric fields. Such coupling can result in valley polarization [8,9], or the valley Hall effect [2] creating a topological current in graphene systems [10] or in 2H-monolayer TMDCs [11]. Specifically, the coupling between a valley pseudospin and an in-plane electric field gives rise to the so-called valley-orbit interaction (VOI) [12,13] suitable for valley manipulation. Based on the VOI mechanism, a unified methodology has been developed to realize functional devices such as valley qubits [13,14], valley FETs [14,15], and valley filters [16].

2H-monolayer TMDCs exhibit novel valley physics [17] distinct from that in graphene. For example, they possess intrinsic inversion symmetry breaking, as opposed to graphene systems that use substrates [18,19] to achieve or vertical electric fields [20-23] to control the breaking. Moreover, while an approximate electron-hole symmetry holds in graphene, the lack of such a symmetry leads to distinct valley physics for holes and electrons [24] in TMDCs. For example, a strong spin-orbit coupling (SOC) exists in the valence bands of TMDCs giving a large spin-orbit splitting in the bands. Due to the time reversal symmetry, both the spins and spin-orbit gaps in top valence bands are opposite in signs for electrons in the two valleys [17]. This is known as the spin-valley coupling and protects holes from inter-valley scattering by non-magnetic impurities thus enhancing the valley coherence of holes.

Previous studies of TMDC-based valleytronics mainly focus on the optical pumping of valley polarization, which exploits the unique opto-valleytronic physics in 2H-monolayer TMDCs, namely, direct band gaps at the two valleys and the valley-dependent selection rule [25]. These properties provide a strong light-matter coupling and allow the generation of valley polarization by application of circularly-polarized light [26-29]. On the other hand, from both scientific and technological perspectives, studies of non-optical approaches should also be essential for TMDC-based valleytronics. In connection to this alternative direction, there have recently been theoretical works proposing the use of ferromagnetic materials [30], line defects [31], or point defect scattering [32] for the generation of valley currents. However, a systematic theoretical study of valley filtering for holes using gated quantum structures is yet to be performed for TMDCs. With such structures it permits valley control via electrical gates and, thus, constitutes a particularly interesting tactic in line with the prevailing practice in the IC industry.

In this work, we study the valley filtering of hole states in lateral quantum structures of 2H-monolayer TMDC systems with electrostatic gating. We start, in **Sec. II**, by deriving a one-band, low-energy effective Hamiltonian from a four-band $k \cdot p$ model, for the topmost valence band in TMDCs. This effective Hamiltonian facilitates the theoretical understanding as well as design of hole-based valleytronic structures. Guided by the Hamiltonian, **Sec. III** presents one-barrier, lateral quantum devices formed of gated structures in TMDCs and lateral sandwich heterostructures such as WS$_2$/MoS$_2$/WS$_2$. These structures can generate valley-dependent hole tunneling and perform the function of valley filtering. In **Sec. IV**, a symmetry-based analysis using the S-matrix is provided for hole transmissions in the quantum structures. **Sec. V** formulates the hole transport problem within a three d-orbital tight binding description of TMDCs, and presents a recursive Green's function (RGF) algorithm for solving hole transmissions in the structures. Specifically, a mixed *r-k* space (i.e., real and momentum space) scheme for the RGF algorithm is developed which reduces the dimensionality of the problem from two to one. In **Sec. VI**, we present numerical results of hole transmissions in the quantum structures. The main findings are the following: i) the tunneling current valley polarization increases with increasing barrier width or height, ii) both the valley-orbit interaction and band structure warping contribute to valley-dependent tunneling, with the former contribution being manifest in structures with asymmetric barriers, and the warping contribution being orientation-dependent and reaching maximum for transmission in the armchair direction, and iii) for transmission ~ 0.1, a current valley polarization of the order of 10% can be achieved. Our study is concluded in **Sec. VII**. The **Appendix** provides a supplement to the discussion of $k \cdot p$ model in **Sec. II**.

## II. K·P EFFECTIVE HAMILTONIAN FOR THE VALENCE BAND

We derive, for 2H-monolayer TMDCs, an effective one-band Hamiltonian for the topmost valence band states near Dirac points, within the $k \cdot p$ theory. Such a Hamiltonian provides insights into various effects in the valley physics of holes in TMDCs.

**Figures 1(a)** and **1(b)** present the top and side views, respectively, of atomic arrangement in a 2H-monolayer TMDC crystal with chemical formula MX$_2$, which is a three-layered structure with one layer of transition metal atoms M = Mo or W sandwiched between two layers of chalcogen atoms X = S, Se, or Te. **Figure 1(c)** shows the corresponding Brillouin zone. Throughout this work, we take the x-axis (y-axis) to be along the armchair (zigzag) direction and the z-axis perpendicular to the xy-plane. With this convention, the wave vectors at K and K' satisfy the relation $\vec{k}_{K'} = -\vec{k}_K$.

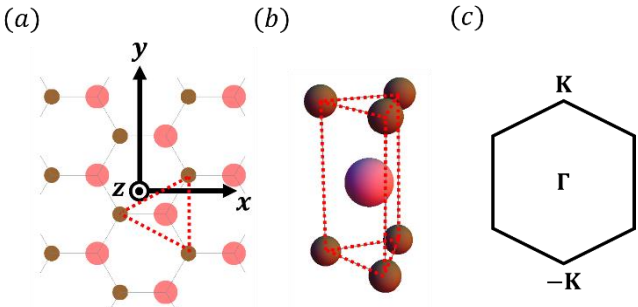

**Figure 1**. (Color online) (a) Top view of 2H-monolayer TMDC in real space, where large pink disks represent metal atoms M and small brown disks represent the chalcogen atoms X. (b) Side view of 2H-monolayer TMDC in real space. (c) The first Brillouin zone.

Main features of the effective Hamiltonian are summarized below. In the presence of an in-plane potential energy variation, the Hamiltonian manifests a valley-dependent term, namely, the valley-orbit interaction (VOI). Other terms that emerge and are valley-dependent include those in association with the presence of band trigonal warping (TW) [33], intrinsic spin-orbit coupling (SOC) [33,34] or Rashba SOC [24,34]. Among these, the TW and VOI terms are found to be dominant and, thus, constitute potential resources for valleytronic applications.

In the following, we will provide a sketch of the theoretical derivation and leave background details to the **Appendix**.

### A. Spinless effective one-band model

We ignore the carrier spin and SOC first. The essential features of valence band structure around Dirac valleys are captured by a four-band $k \cdot p$ Hamiltonian [33], which is constructed in the basis of the Bloch functions at K (K') with an even parity in the z-direction, e.g., $\{\Psi^v_{A_1(A_1)}, \Psi^{c+2}_{E_{1-}(E_{1+})}, \Psi^c_{E_{1+}(E_{1-})}, \Psi^{v-3}_{E_{1-}(E_{1+})}\}$. Here, the superscript denotes the corresponding energy band to which the state belongs, with c (v) standing for the first conduction (valence) band and c+2 (v-3) the third conduction (fourth valence) band; the subscript indicates the corresponding irreducible group representation to which the state belongs, of $C_{3h}$, the group of wave vectors at K (K'). The Hamiltonian in the four-band $k \cdot p$ model reads

$$H_{4b} = \begin{pmatrix} E^v + V_{ext} & \tau P_1 k_+ & \tau P_2 k_- & \tau P_3 k_+ \\ \tau P_1^* k_- & E^{c+2} + V_{ext} & \tau P_4 k_+ & 0 \\ \tau P_2^* k_+ & \tau P_4^* k_- & E^c + V_{ext} & \tau P_5 k_- \\ \tau P_3^* k_- & 0 & \tau P_5^* k_+ & E^{v-3} + V_{ext} \end{pmatrix} \quad (1)$$

, where $\tau = 1 (-1)$ for K (K') valley, $k_\pm = \partial_y \pm i\tau \partial_x$, $V_{ext} = V_{ext}(x, y)$ is the in-plane potential energy, $E^b$ ($b = v, c+2, c,$ or $v-3$) is the band edge energy at K or K', and $P_\mu$ ($\mu = 1 \sim 5$) is, except for a trivial prefactor, the momentum matrix element between corresponding basis states. Note that $P_\mu$'s are real-valued [33]. We employ the Löwdin perturbation theory [35] and treat the off-diagonal $k \cdot p$ terms and diagonal $V_{ext}$ as perturbations to the third order. This gives, for low-energy holes in the topmost valence band, the following one-band electron Hamiltonian

$$H_{1b}(\tau) = H_0 + H_{TW} + H_{VOI},$$

$$H_0 = \frac{\hbar^2 k^2}{2m^*} + V_{ext},$$

$$H_{TW} = \tau\alpha_{TW}(k_y^3 - 3k_y k_x^2),$$
$$H_{VOI} = \tau\alpha_{VOI}(\nabla V_{ext} \times \vec{k}) \cdot \hat{z}, \quad (2)$$

where

$$\alpha_{TW} = \frac{2P_1 P_2 P_4}{(E^v - E^{c+2})(E^v - E^c)} + \frac{2P_2 P_3 P_5}{(E^v - E^{v-3})(E^v - E^c)} \quad (2a)$$

$$\alpha_{VOI} = -\frac{P_1^2}{2(E^v - E^{c+2})^2} - \frac{P_2^2}{2(E^v - E^c)^2} - \frac{P_3^2}{2(E^v - E^{v-3})^2} \quad (2b)$$

and

$$m^* = \frac{\hbar^2}{2}\left(\frac{P_1^2}{E^v - E^{c+2}} + \frac{P_2^2}{E^v - E^c} + \frac{P_3^2}{E^v - E^{v-3}}\right)^{-1} \quad (2c)$$

is the effective mass. Two $\tau$-dependent terms enter the effective Hamiltonian, namely, $H_{TW}$ corresponding to the band warping and $H_{VOI}$ corresponding to the VOI. As $H_{VOI}$ comes from couplings of the top valence band to the rest in the model, it consists of corresponding contributions that scale, respectively, inversely with $(E^v - E^{c+2})^2$, $(E^v - E^c)^2$, and $(E^v - E^{v-3})^2$. In comparison to the corresponding expression of $\alpha_{VOI}$ in gapped graphene, where it scales inversely with $(E^v - E^c)^2$, $\alpha_{VOI}$ here is weaker due to the typically large gap size contrast between the two materials.

Overall, the spinless effective Hamiltonian in Eq. (2) is sufficient to capture the main valley-dependent effects and, thereby, can serve as a useful guidance for the design of valleytronic devices. The inclusion of spin and SOC in the model slightly modifies the parameters $m^*$, $\alpha_{TW}$ and $\alpha_{VOI}$, as will be explained in the next subsection.

### B. Effects of spin-orbit coupling

The effects of SOC are described in more details in the **Appendix**. Basically, they can be classified into two categories, namely, intrinsic and extrinsic ones.

The intrinsic effects derive from the coupling of electron spins to the crystal potential and modify the band structure obtained in the spinless model. Overall, the inclusion of intrinsic SOC effects produces a shift of the various band edges by opening up the spin-orbit gaps, resulting in a relative band edge shift of the order of $\Delta_{so}/\Delta_{BG}$, where $\Delta_{so}$ is the typical spin-orbit gap and $\Delta_{BG}$ the typical band gap in TMDCs. Using Eqs. (2a) and (2c), this produces in $m^*$ and $\alpha_{TW}$ a corresponding relative change of the order of $\Delta_{so}/\Delta_{BG}$. Using $\Delta_{so} = O(100 meV)$ and $\Delta_{BG} = O(1eV)$, such a change is typically small. If desired, such a change can be incorporated by re-defining the various band edges in the spinless model.

On the other hand, there are also extrinsic SOC effects coming from the coupling between spins and external electric fields. These include the Rashba SOC in the presence of a vertical electric field and the SOC-induced change in VOI. As discussed in the **Appendix**, if we take the in-plane and out-of-plane electric fields to be of the same order, then the Rashba SOC is a factor of $\Delta_{so}/\Delta_{BG}$ smaller than $H_{VOI}$ given in Eq. (2). For this reason, we will ignore the Rashba SOC in our study of the valley-dependent transport. As for the SOC-induced change in $\alpha_{VOI}$, it is estimated to be of the order of $\Delta_{so}/\Delta_{BG}$ just as that in the case of $m^*$ and $\alpha_{TW}$.

Overall, the above discussion confirms the applicability of the effective Hamiltonian given by Eq. (2) to hole-based valleytronics in TMDCs.

## III. QUANTUM STRUCTURES FOR VALLEY FILTERING

We now explore electrical valley filtering of holes in TMDCs going through a lateral quantum barrier. In the corresponding tunneling problem to be considered here, the carrier energy E and transverse wave vector $k_t = k_x(k_y)$ for tunneling in the zigzag (armchair) direction are taken to be conserved quantities. In addition, the valley index $\tau$ is also a constant of the tunneling due to the spin-valley locking caused suppression of inter-valley scattering.

Depending on the orientation of the barrier, two cases are analyzed as follows.

### A. Barrier stripe along the armchair direction

First, we consider the case where the barrier stripe lies along the armchair direction, i.e. $V_{ext} = V_{ext}(y)$ (see **Figure 2(a)**). Using Eq. (2), we write the effective Hamiltonian for a given $k_x$

$$H_{AC}(k_y \to -i\partial_y, y; k_x) = H_{kinetic} + H_{potential} + \frac{\hbar^2 k_x^2}{2m^*}$$
$$H_{kinetic}(k_y \to -i\partial_y; k_x) = \frac{\hbar^2 k_y^2}{2m^*} + \tau\alpha_{TW}k_y^3 - \tau 3\alpha_{TW}k_x^2 k_y$$
$$H_{potential}(y; k_x) = V_{ext}(y) - \tau\alpha_{VOI}k_x\partial_y V_{ext}(y) \quad (3)$$

Under the specific classification of terms in Eq. (3), $H_{VOI}$ generates a valley-dependent term "$-\tau\alpha_{VOI}k_x\partial_y V_{ext}(y)$" in the potential energy $H_{potential}$ that can result in valley-dependent tunneling. We note that in systems with a homogeneous band gap, the effect of $H_{VOI}$ is expected to be significantly suppressed. This can be understood in terms of the Ehrenfest theorem, namely, if we treat "$-\tau\alpha_{VOI}k_x\partial_y V_{ext}(y)$" in Eq. (3) as a perturbation and perform the first-order perturbation-theoretical estimation, then we obtain $-\tau\alpha_{VOI}k_x <\partial_y V_{ext}(y)> \propto d<p_y>/dt = 0$ within the effective one-band approximation. Two structures are available in order to dodge the Ehrenfest theorem-caused suppression. For practical applications, let us focus on a hole energy which on the incident side is near the local valence band edge. First, one could use a suitably high barrier such that in the barrier region the hole energy would be deep in the band gap and sufficiently away from the local valence band edge. This would invalidate the low-energy, one-band approximation in the barrier region and hence the Ehrenfest theorem valid for the low-energy regime. Second, one could alternatively use a lateral heterostructure. In this case the various coefficients in the one-band Hamiltonian would become material-dependent and hence vary in space. Therefore, a straightforward application of the Ehrenfest theorem would fail. With either structure, the existence of a sizable, non-vanishing $<-\tau\alpha_{VOI}k_x\partial_y V_{ext}(y)>$ can be made possible. As such, we identify the presence of either a band gap inhomogeneity or a sufficiently high barrier in the structure as the necessary condition for a clear manifestation of the VOI-induced valley dependent tunneling. Such a condition will be numerically verified in **Sec. VI**.

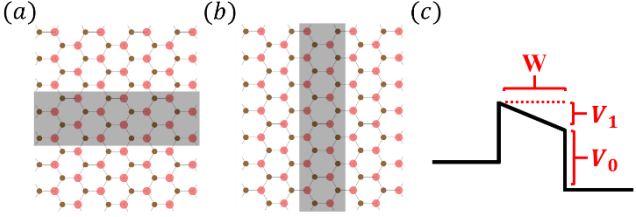

**Figure 2**. (Color online) (a) A one-barrier structure with the gray area indicating a barrier lying along the armchair direction. (b) A one-barrier structure with the gray area showing a barrier lying along the zigzag direction. (c) The inverted valence band edge diagram of a one-barrier structure with the barrier height $V_0$ and the source-drain bias $V_1 = 0$ ( $V_1 \neq 0$ ) for a symmetric (asymmetric) structure.

Eq. (3) indicates that $H_{TW}$ also contributes to the kinetic energy, $H_{kinetic}$, a valley-dependent term " $\tau\alpha_{TW}k_y^3 -\tau 3\alpha_{TW}k_x^2 k_y$ " that is odd in $k_y$. Naively, one would think that such a term could be effective in generating a valley-polarized current. However, their effect on tunneling is negligible, as explained in the following. We consider a tunneling electron with energy $E$ lower than the barrier height. For such an electron, the corresponding wave vector $k_y$ is complex with a finite imaginary part describing the attenuation of wave amplitude in the barrier. We analyze the valley dependence of $k_y$ in the barrier by analytically continuing the Hamiltonian $H_{AC}$ into the complex $k_y$-plane. Using $H_{kinetic}$ in Eq. (3), the local valence band dispersion in the barrier can be written as

$$E(k_y, \tau) = \sum_{i=0,1,2...} \tau^i a^i k_y^i \qquad (3')$$

, where $a^i$ are real-valued constants. For an incident carrier at the given energy E, the above equation implies multiple solutions of wave vectors {$k_y$'s}, with each constituting a channel available to the carrier tunneling. In the following, we show that {$k_y$'s} have identical distributions in the imaginary part for $\tau = \pm 1$. Let $k_{y,K}(E)$ be a solution at energy E for $\tau = 1$. Then it follows that { $k_{y,K}(E)$ , $k_{y,K}^*(E)$ } are both solutions, as is well known for an equation such as Eq. (3′) with real-valued constants. On the other hand, it can be verified that { $-k_{y,K}(E)$ , $-k_{y,K}^*(E)$ } are solutions for $\tau = -1$. Now we compare the solutions for $\tau = \pm 1$. We see that the imaginary parts of { $k_{y,K}(E)$ , $k_{y,K}^*(E)$ } and those of { $-k_{y,K}(E)$ , $-k_{y,K}^*(E)$ } are identical. This suggests the same tunneling rates for carriers of opposite valleys. Thereby, as opposed to $H_{VOI}$ , $H_{TW}$ does not effectively lead to valley-dependent tunneling.

### B. Barrier stripe along the zigzag direction

Next, we consider the case where the barrier lies along the zigzag direction, i.e. $V_{ext} = V_{ext}(x)$ (see **Figure 2(b)**). The effective Hamiltonian becomes

$$H_{ZZ}(k_x \to i\partial_x, x; k_y) = H_{kinetic} + H_{potential} + \frac{\hbar^2 k_y^2}{2m^*}$$

$$H_{kinetic}(k_x \to i\partial_x; k_y) = \frac{\hbar^2 k_x^2}{2m^*} - \tau 3\alpha_{TW} k_y k_x^2$$

$$H_{potential}(x; k_y) = \tau\alpha_{TW} k_y^3 + V_{ext}(x) + \tau\alpha_{VOI} k_y \partial_x V_{ext}(x)$$

(4)

We note that as argued in the previous case, $H_{VOI}$ can generate a valley-dependent tunneling when the band gap varies in space. On the other hand, as opposed to the first case, $H_{TW}$ now contributes to the kinetic energy a term " $-\tau 3\alpha_{TW} k_y k_x^2$ " that is quadratic in $k_x$. Such a term can modify the carrier effective mass in a valley-dependent way. Therefore, when the Hamiltonian is analytically continued into the tunneling regime, it results, due to the mass difference between carriers of opposite valleys, in a valley-contrasted tunneling behavior where carriers of one valley have a larger mass and weaker tunneling than those of the other valley. By comparing the two cases of barrier orientations, we find that $H_{TW}$ shows a strong anisotropy in producing the valley contrast.

For further investigation of valley-dependent hole transport below, we shall consider barriers with symmetric ( $V_1 = 0$ ) or asymmetric ( $V_1 \neq 0$ ) profiles as shown in **Figure 2(c)**, in structures with homogeneous or inhomogeneous band gaps. We define the valley polarization of a tunneling current as $P_v \equiv (T_K - T_{-K})/(T_K + T_{-K})$ , with $T_{K(-K)}$ referring to the transmission probability of hole states at the K (K') valley, and study $P_v$ with a symmetry-based analysis in **Sec. IV** and numerically in **Sec. VI**.

### IV. SYMMETRY-BASED ANALYSIS

The symmetry-based analysis is performed within the formalism of S-matrix. **Figure 3** shows the scattering of a hole off a barrier, where $\Psi_{E,k_t,\tau}^{\xi,in(out)}$ is the wave in Region-$\xi$ ( $\xi$ = I or II) moving toward (away from) the barrier. The S-matrix relates the incoming current amplitudes A (carried by $\Psi^{I,in}_{E,k_t,\tau}$) and D (carried by $\Psi^{II,in}_{E,k_t,\tau}$) to the outgoing ones B (carried by $\Psi^{I,out}_{E,k_t,\tau}$) and C (carried by $\Psi^{II,out}_{E,k_t,\tau}$):

$$\begin{pmatrix} B \\ C \end{pmatrix} = S_{E,k_t,\tau} \begin{pmatrix} A \\ D \end{pmatrix} = \begin{pmatrix} r_{E,k_t,\tau} & t'_{E,k_t,\tau} \\ t_{E,k_t,\tau} & r'_{E,k_t,\tau} \end{pmatrix} \begin{pmatrix} A \\ D \end{pmatrix} \quad (5)$$

with the matrix elements $r_{E,k_t,\tau}$ and $r'_{E,k_t,\tau}$ being the reflection and $t_{E,k_t,\tau}$ $t'_{E,k_t,\tau}$ the transmission amplitudes. In particular, for a hole moving from Region-I to Region-II, the transmission probability is given by $T_{E,k_t,\tau} = |t_{E,k_t,\tau}|^2$.

For a given set of E, $k_t$ and $\tau$, the S-matrix given in Eq. (5) is 2 x 2. Additional restrictions on the S-matrix arise from the probability conservation and also the symmetry in the system, as discussed next.

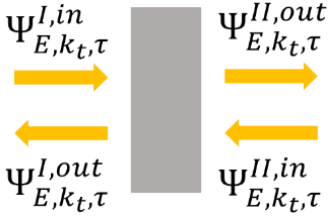

**Figure 3**. (Color online) The schematic diagram showing incident and reflected waves in the two regions labelled as I and II of the one-barrier structure.

### A. Probability conservation

As is well known, due to the probability conservation law, $S_{E,k_t,\tau}$ is required to be unitary. This leads to

$$|t_{E,k_t,\tau}| = |t'_{E,k_t,\tau}|. \quad (6)$$

### B. Time reversal symmetry

Under the time reversal operation, $\Psi^{\xi,in(out)}_{E,k_t,\tau}$ transforms into $\Psi^{\xi,out(in)}_{E,-k_t,-\tau}$ and, correspondingly, Eq. (5) is transformed into

$$\begin{pmatrix} A^* \\ D^* \end{pmatrix} = S_{E,-k_t,-\tau} \begin{pmatrix} B^* \\ C^* \end{pmatrix} \quad \text{or} \quad \begin{pmatrix} B \\ C \end{pmatrix} = S^T_{E,-k_t,-\tau} \begin{pmatrix} A \\ D \end{pmatrix}.$$
(5′)

using the unitarity of $S_{E,-k_t,-\tau}$. Comparing Eq. (5′) to Eq. (5), we obtain $S^T_{E,-k_t,-\tau} = S_{E,k_t,\tau}$ and, thus, $|t_{E,k_t,\tau}| = |t'_{E,-k_t,-\tau}|$. Combined with Eq. (6), it leads to

$$|t_{E,k_t,\tau}| = |t_{E,-k_t,-\tau}| \quad (7)$$

Eq. (7) implies the following. For normal incidence, where $k_t = 0$, Eq. (7) concludes a vanishing tunneling current valley polarization in all cases independent of the barrier shape and orientation. In other words, the analysis here identifies $k_t \neq 0$ as a necessary condition on incident carriers for the generation of tunneling current valley polarization.

### C. Reflection symmetry

For barriers lying along the zigzag direction, the reflection symmetry under the transformation x → -x is always broken due to the dissimilarity between M and X. However, for barriers lying along the armchair direction, the situation varies depending on $V_1$.

In the case where $V_1 = 0$, the structure exhibits the reflection symmetry under the transformation y → -y. Under the reflection, $\Psi^{\xi,in(out)}_{E,k_x,\tau}$ transforms into $\Psi^{\xi,out(in)}_{E,k_x,-\tau}$ and, correspondingly, Eq. (5) becomes

$$\begin{pmatrix} A \\ D \end{pmatrix} = S_{E,k_x,-\tau} \begin{pmatrix} B \\ C \end{pmatrix} \quad \text{or} \quad \begin{pmatrix} B \\ C \end{pmatrix} = S^\dagger_{E,k_x,-\tau} \begin{pmatrix} A \\ D \end{pmatrix} \quad (5'')$$

Comparing Eq. (5″) to Eq. (5), we obtain $S^\dagger_{E,k_x,-\tau} = S_{E,k_x,\tau}$ or $|t_{E,k_x,\tau}| = |t'_{E,k_x,-\tau}|$. Combined with Eq. (6), it gives

$$|t_{E,k_x,\tau}| = |t_{E,k_x,-\tau}| \quad (8)$$

Eq. (8) concludes a vanishing tunneling current valley polarization for any incident angle, when $V_1 = 0$. Therefore, for one-barrier structures with barriers along the armchair direction, $V_1 \neq 0$ is another necessary condition, apart from $k_t \neq 0$, on the generation of tunneling current valley polarization.

The result of this section along with that derived in **Sec. III** within the effective Hamiltonian-based analysis is summarized in **Table I** below.

| Barrier Orientation ($k_t \neq 0$) | Symmetric Barrier | Asymmetric Barrier |
|---|---|---|
| Armchair | $P_v = 0$ | VOI* |
| Zigzag | TW | TW, VOI* |

**Table I**. TW- and VOI- based valley-polarized hole tunneling in various situations. All cases require oblique incidence ($k_t \neq 0$) for $P_v$ to be non-vanishing. "*" here indicates the condition of either a band gap inhomogeneity or a suitably high barrier is further required for a sizable polarization.

As summed up in **Table I**, in all of the cases the condition

$k_t \neq 0$ is required for a non-vanishing tunneling current valley polarization. We note that this condition can be experimentally realized by, for example, sampling the transmitted current [6] or generating the incident current [36,37], at an angle with respect to the barrier's normal direction.

## V. RECURSIVE GREEN'S FUNCTION APPROACH

**Sec. V-A** presents a generic, efficient recursive Green's function method in the mixed *r-k* space, for the study of carrier transmission in 2D material based, lateral quantum structures. **Sec. V-B** discusses the application of the method to TMDC systems.

### A. Mixed *r-k* space formulation for recursive Green's function

**Figure 4(a)** shows a quantum structure with lattice translational symmetry in the x-direction and a potential variation in the y-direction generated with electrostatic gating, for example, on top of the gray area. We consider electron transport in the y-direction in the tight-binding model. The system is taken to have a square lattice structure. For illustration, only nearest-neighbor hoppings $\{t_+, t_-, u_+, u_-\}$ and next-nearest-neighbor hoppings $\{v_+, v_-, w_+, w_-\}$ are included in the model. In the case of TMDCs, a supercell can be chosen to transform the hexagonal lattice of TMDCs into a square lattice, as will be shown in **Sec. V-B**.

The wave equation in the tight-binding model is given by

$$H_j C_{i,j} + u_+ C_{i+1,j} + u_- C_{i-1,j} + t_+ C_{i,j+1} + t_- C_{i,j-1} + v_+ C_{i-1,j+1}$$
$$+ v_- C_{i+1,j-1} + w_+ C_{i+1,j+1} + w_- C_{i-1,j-1} = EC_{i,j}, \quad (9)$$

where $C_{i,j}$ is the amplitude, which can be multi-component, at the cell with coordinates (i,j) and $H_j$ is the corresponding on-site part of the Hamiltonian. Let $k_x$ be the transverse Bloch wave vector and $a_x$ the lattice constant in the x-direction. Using the Bloch theorem, e.g., $C_{i+1,j} = e^{ik_x a_x} C_{i,j}$, we can effectively remove the x-dependence from Eq. (9) and reduce it to

$$\tilde{H}_j(k_x) C_{i,j} + \tilde{t}_+(k_x) C_{i,j+1} + \tilde{t}_-(k_x) C_{i,j-1} = EC_{i,j}. \quad (9')$$

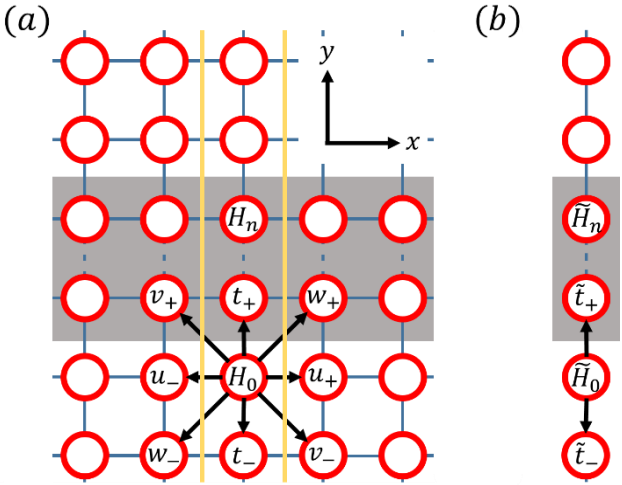

**Figure 4.** (Color online) (a) A lateral quantum structure in two-dimensional square lattice. Unit cells are represented by circles. The gray area indicates a barrier. The region bounded by vertical yellow lines denotes a column of unit cells. $H_j$ (j = integer) is the on-site part of the tight-binding Hamiltonian for a cell with coordinates (i,j). $t_\pm$, $u_\pm$, $v_\pm$, and $w_\pm$ are hopping matrices. (b) Reduction of the two-dimensional lattice to a one-dimensional chain in the mixed *r-k* space, with $\tilde{H}_j$ and $\tilde{t}_\pm$ being, respectively, the effective on-site cell Hamiltonian and nearest-neighbor hoppings.

With the x-coordinate fixed at "*i*", Eq. (9′) now involves only a single column of unit cells. **Figure 4(b)** shows the corresponding reduced system - a one-dimensional chain. The on-site part of the Hamiltonian ($\tilde{H}_j$) and nearest-neighbor hoppings ($\tilde{t}_\pm$) in the chain are given by

$$\tilde{H}_j(k_x) = H_j + u_+ e^{ik_x a_x} + u_- e^{-ik_x a_x},$$
$$\tilde{t}_+(k_x) = t_+ + w_+ e^{ik_x a_x} + v_+ e^{-ik_x a_x},$$
$$\tilde{t}_-(k_x) = t_- + v_- e^{ik_x a_x} + w_- e^{-ik_x a_x}. \quad (10)$$

The remaining task is then to apply the standard recursive Green's function technique [38] and calculate the transmission coefficient in the effective chain.

We make two notes about the present method. First, in an alternative recursive Green's function approach, one could perform the calculation in the xy-space, where one replaces the structure with an infinite x-dimension with one of a finite x-dimension, e.g., a nanoribbon. In order to avoid edge effects as well as finite size effects, this nanoribbon would have to be wide enough. The present approach in mixed $k_x$ and y space is obviously much more efficient from a computational point of view, and naturally permits the study of transport quantities such as the transmission coefficient as a function of the transverse wave vector, $k_t$, in the case of lateral quantum structures. Second, as the present approach solves the wave amplitude on each atomic site, it automatically satisfies the requirement of current continuity across the interface between different regions, as opposed to bulk band structure-based approaches [16,39], where bulk solutions are firstly obtained in each region of the quantum structure and then matched across the interface. For such approaches, special care has to be taken to ensure current continuity at the interface.

### B. TMDC-based quantum structures

We now apply the method developed in **Sec. V-A** to TMDC-based lateral quantum structures.

For a semi-quantitative study, we adopt a minimal, three-orbital tight-binding description that uses

$d_{z^2}, d_{x^2-y^2}, d_{xy}$ of the metal atom M as the basis states, which are known to be the dominant constituent orbitals near the conduction and valence band edges [40]. Overall, since the $D_{3h}$ point-group symmetry of TMDCs is faithfully retained, we expect that the omission in the model of the *p* orbitals from the chalcogen atom X, even though they also contribute to the band edge states, would not pose any essential problem for a semi-quantitative study.

A few more notes are given below about the tight-binding model. First, in order to describe the warping and effective mass of the valence bands with a reasonable accuracy, hopping terms up to the third-nearest neighbors are included (see **Figure 5(a)**). Second, following our earlier discussion in **Sec. II-B** about the SOC, the model accounts for the dominant SOC effect – the resultant spin-orbit gap by including it in the on-site orbital energy, and ignores the less significant Rashba SOC. In fact, the Rashba SOC is automatically excluded from the model. With the coupling being proportional to the matrix element of *z* [41] between states, it vanishes between any pair of states from $\{d_{z^2}, d_{x^2-y^2}, d_{xy}\}$ where all the states are of even parity in the z-direction. Third, as a main contribution to the VOI comes, according to Eq. (2b), from the coupling between the first conduction band and the first valence band, the model is able to accommodate this interaction. Therefore, overall, the model is able to capture major valley-dependent effects that are of interest to our investigation. If desired, one can use the three-orbital model as an effective model and adjust the various on-site and hopping parameters to fit it with an extensive one and improve the quantitative aspect of the model.

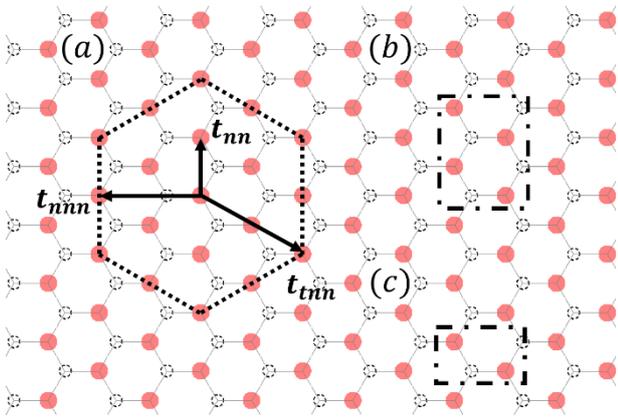

**Figure 5.** The chalcogen atoms are denoted by the dashed circle, and their atomic orbitals are ignored in the three-orbital model. (a) The hoping terms considered in the calculation, with $t_{nn}$ the nearest neighbor hopping, $t_{nnn}$ the next-nearest neighbor hopping, and $t_{tnn}$ the third-nearest neighbor hopping. (b) Supercell for barriers along the armchair direction. (c) Supercell for barriers along the zigzag direction.

The formulation of **Sec. V-A** is developed for the structure of a square lattice. In order to apply it to the TMDC system, we choose the supercells as shown in **Figure 5(b)** for barriers along the armchair direction and **Figure 5(c)** for barriers along the zigzag direction. It can be verified that in each case a unit cell is "nearest-neighbor" coupled, on the supercell scale, to the eight neighboring cells, among which four of them share common edges and the rest share common vertices with it. It can be verified that this inter-supercell coupling does not extend beyond nearest neighbors.

## VI. NUMERICAL RESULTS

We apply the formalism presented in **Sec. V** to the numerical study of valley filtering of holes in TMDC lateral quantum structures, with the tight binding parameters and the band offsets between different TMDCs in the study taken from references 40 and 42.

Our numerical results of hole transmissions in various one-barrier structures are presented in **Figures 6-8**. In **Figures 6** and **7**, we consider homogeneous structures with barriers lying along the armchair and zigzag directions, respectively, while **Figure 8** compares homogeneous with heterogeneous structures. In all figures, we take the valence band edge of the structure on the incidence side to be the zero reference energy.

We summarize below the common features in **Figures 6** and **7**. **Figures 6(a)** and **7(a)** present valley-averaged hole transmission coefficients, showing a magnitude which is near unity for hole energy above the barrier and rapidly decreases for hole energy below the barrier. **Figures 6(b)** and **7(b)** present corresponding tunneling current valley polarizations. We see that they vanish at $k_t = 0$ and increase with the magnitude of the transverse wave vector $k_t$, in agreement with **Table I**. Moreover, the valley-dependent tunneling as a mechanism of valley filtering is verified. As shown in the figures, outside the tunneling regime, e.g., when transmissions ∼ 1, the resultant tunneling current valley polarization becomes insignificantly small. On the other hand, the polarization rises up as the carrier energy moves into the tunneling regime, with the resultant polarization depending on the barrier width as well as the height - the wider or higher the barrier, the larger the polarization. This is evident in **Figure 6(c)** when comparing polarizations in the two cases where $V_1$ is fixed at 0.01 eV. The polarization increases with increasing $V_0$. A similar trend also holds in **Figure 7(c)** when comparing the case where $V_0 = 0.01$ eV, W = 100 Å to either that where $V_0$ is increased to 0.02 eV or that where W is increased to 200 Å. **Figures 6(d)** and **7(d)** compare tunneling current valley polarizations for $MoSe_2$ and $WSe_2$. Within the three d-orbital tight-binding model employed here, the two materials show analogous behaviors, which can be attributed to their identical crystal structures and similar band structures. We have also performed calculations for $MoS_2$ and $WS_2$, and found, due to the dominance of valence bands by the d orbitals of metallic ions, that the resultant polarization curves for $MoS_2$ and $WS_2$ are, respectively, nearly identical to those of $MoSe_2$ and $WSe_2$. Last, in **Figures 6(c)-(d)** and **7(c)-(d)**, a phenomenon of oscillations in $P_v$ is noticeable. These oscillations also show up in **Figures 6(a)** and **7(a)** and are attributed to the Fabry-Pérot type resonance occurring between the two barrier-electrode interfaces.

Next, we discuss the contrast between **Figures 6** and **7**. This contrast is primarily manifested as an order-of-magnitude difference in $P_v$'s with, for example, $P_v \approx O(0.001\%)$ in **Figure 6(c)** and $P_v \approx O(10\%)$ in **Figure 7(c)**. According to **Table 1**, the dominant mechanisms of tunneling current valley polarization are different in the two cases – while it is VOI-based in **Figure 6(c)**, it is TW-based in **Figure 7(c)**. In the case of **Figure 6(c)**, because of the Ehrenfest theorem, the VOI mechanism is greatly suppressed. Therefore, it results, for the total carrier energy, in one-barrier structures with barriers lying along the armchair direction. Results in (a)-(c) are calculated using the parameters of $WSe_2$. (a) Valley-averaged transmissions for various $k_x$'s with barrier height $V_0 = 0.01\,eV$, source-drain bias $V_1 = 0.01\,eV$ and barrier width $W$ nearly 100 Å. (b) Tunneling current valley polarizations for various $k_x$'s in the same structure considered in (a). (c) Tunneling current valley polarization for various barrier heights ($V_0$) and biases ($V_1$) at $k_x = 0.05 a^{-1}$. (d) Tunneling current valley polarization for two TMDCs with $V_0 = 0.01\,eV$, $V_1 = 0.01\,eV$, $W \approx 100$ Å and $k_x = 0.05 a^{-1}$.

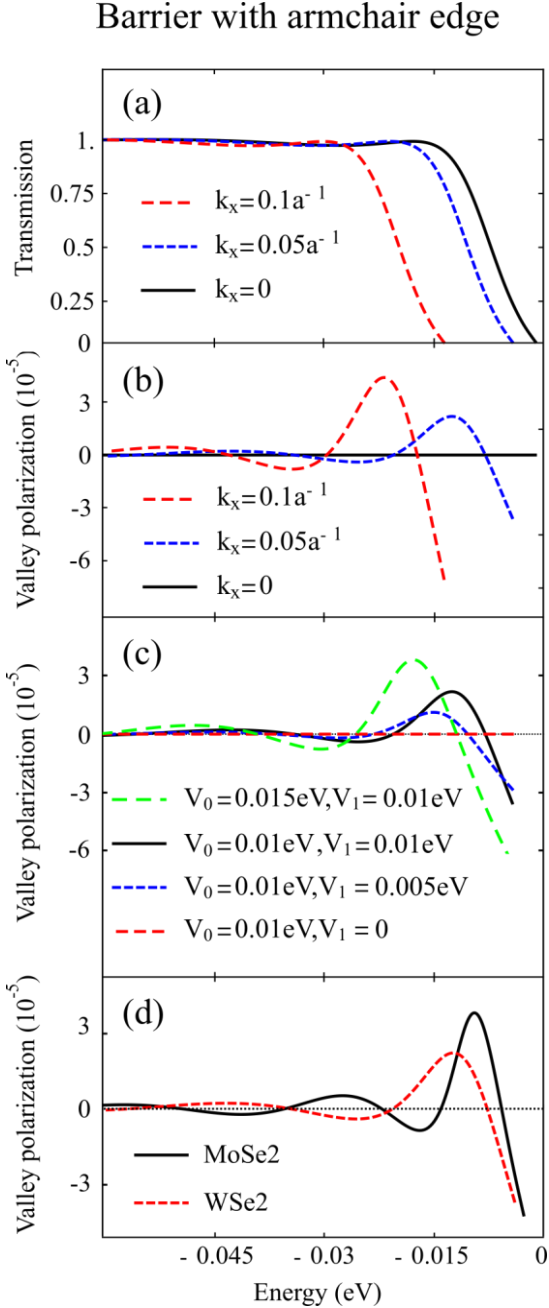

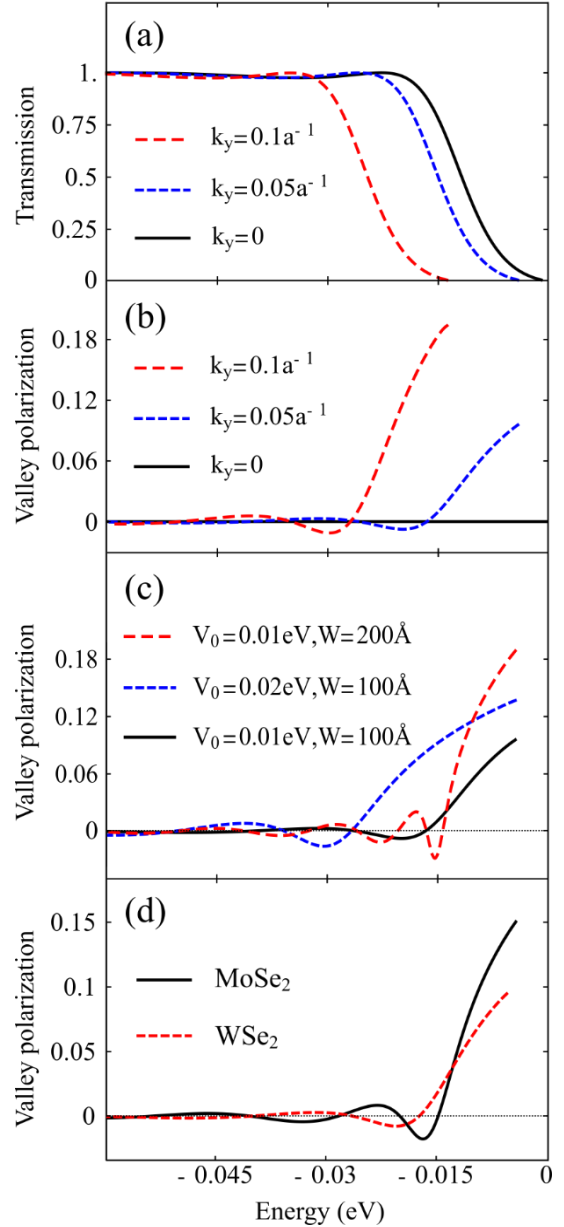

**Figure 6.** (Color online) Hole transmissions and corresponding tunneling current valley polarizations, i.e., $P_v$, as functions of

**Figure 7.** (Color online) Hole transmissions and corresponding tunneling current valley polarizations, i.e., $P_v$, as functions of total carrier energy, in symmetric, one-barrier structures with barriers lying along the zigzag direction and the source-drain bias

$V_1 = 0\ eV$. For (a), (b), and (c), the results are obtained with the tight-binding parameters of $WSe_2$. (a) Valley-averaged transmissions for various transverse wave vectors $k_y$'s with barrier height $V_0 = 0.01\ eV$ and width $W$ nearly $100$ Å. $a$ = lattice constant. (b) Tunneling current valley polarizations for various $k_y$'s in the same structure considered in (a). (c) Tunneling current valley polarizations for various barrier heights ($V_0$) and widths ($W$) at $k_y = 0.05a^{-1}$. (d) Tunneling current valley polarizations for two TMDCs with $V_0 = 0.01 eV$, $V_1 = 0 eV$, $W \approx 100$ Å and $k_y = 0.05 a^{-1}$.

homogeneous structures considered here, in a strong orientational dependence of tunneling current valley polarization.

The weak strength of VOI effect illustrated in **Figure 6** can be improved in several ways. For example, in **Figure 6(c)**, when comparing the three curves with $V_0$ all fixed at 0.01 eV, it shows that the polarization increases with increasing $V_1$. Alternatively, one can invoke a violation of the Ehrenfest theorem by introducing a suitably high barrier or a band gap inhomogeneity such as that in a heterostructure, and reinstate the VOI effect, as discussed below.

In **Figure 8**, we study tunneling current valley polarizations in heterostructures. Specifically, we consider $WS_2 / MoS_2 / WS_2$ [43,44] with the valence band offset taken to be 0.45 eV between $WS_2$ and $MoS_2$. We compare the heterostructure to the homogeneous structure of $MoS_2$ with a barrier height matching the offset. **Figure 8(a)** presents the case where the barrier is oriented in the armchair direction. It shows that the VOI-based polarization of about the order of 1% is generated, with the heterostructure having the higher polarization than the homostructure. On the other hand, **Figure 8(b)** presents the case where the barrier stripe is oriented in the zigzag direction, and it shows that the homostructure gives rise to the higher polarization. Moreover, we find that when $V_1$ is turned on the polarization is reduced. We interpret this reduction as an indication of the VOI contribution being opposite in sign to that of the TW - the dominant contribution at $V_1 = 0$. Using the amount of reduction as an estimate, we obtain from the graph that the VOI-contributed $P_v$ is around 2% in the homogeneous case and 5% in the heterogeneous case. A comparison between the VOI contributions in **Figures 8(a)** and **8(b)** shows that the VOI effect varies relatively slowly in order of magnitude with the barrier orientation.

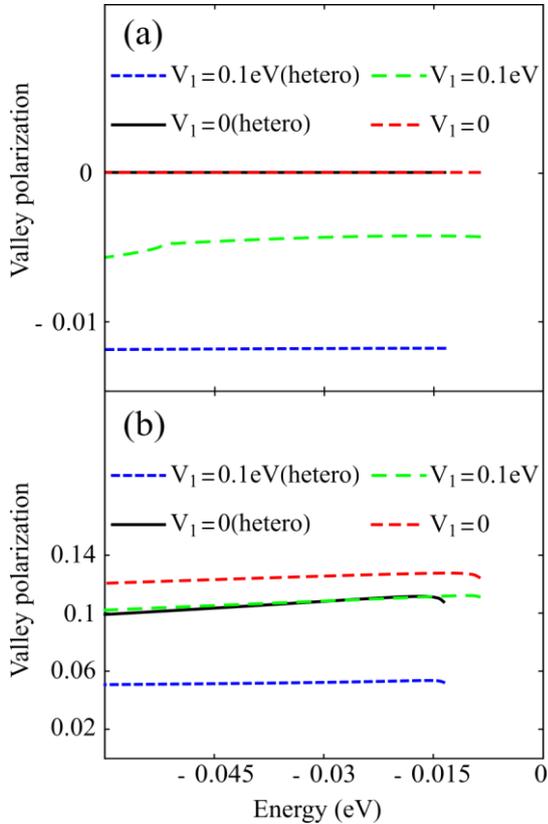

**Figure 8.** (Color online) The comparison of tunneling current valley polarization between a $WS_2 / MoS_2 / WS_2$ heterostructure and a homogeneous $MoS_2$ quantum structure with the same barrier width and height. Curves labelled "hetero" refer to the tunneling current valley polarization in the heterostructure and those without the label refer to that in the homogeneous structure. (a) Tunneling current valley polarization for barriers along the armchair direction with $V_0 = 0.45\ eV$, $W \approx 10$ Å and $k_x = 0.1 a^{-1}$, for $V_1 = 0\ eV$ or $0.1\ eV$. (b) Tunneling current valley polarization for barriers along the zigzag direction with $V_0 = 0.45 eV$, $W \approx 10$ Å and $k_y = 0.1 a^{-1}$, for $V_1 = 0\ eV$ or $0.1\ eV$. In (a), since the VOI effect is dominant, the polarization vanishes at $V_1 = 0\ eV$.

## VII. CONCLUSION

In summary, we have derived an effective one-band $k \cdot p$ Hamiltonian for the valence bands, which can capture main valley-dependent effects and can be used to explore electro-valleytronics of holes in monolayer TMDCs. Using this Hamiltonian as a guidance, an electrical valley filtering structure has been proposed, which is formed of a lateral quantum structure with a potential barrier, with the filtering based on valley-dependent tunneling rates. The valley filtering physics in such a structure has been investigated with both a S matrix-based analysis and a numerical calculation of transmission coefficients with recursive Green's function method.

Generally, we find that the tunneling current valley polarization increases with the barrier width and height. Specifically, two effects on valley filtering are identified. The effect of trigonal warping is shown to be strongly orientation-dependent, being minimal in structures with armchair-oriented barriers and maximal in those with zigzag-oriented ones. In addition to this effect, in structures with asymmetric barriers, a VOI mechanism emerges contributing to valley filtering in a relatively isotropic fashion. For homostructures, the VOI contributed tunneling current valley polarization is largely suppressed. However, by introducing either a band gap variation or a suitably high barrier into the structure, the VOI effect can be significantly restored.

Overall, our study demonstrates that for transmission ~ 0.1 a tunneling current valley polarization ~ 1-10% can be achieved via electrical gate control in lateral, TMDC-based one-barrier structures. Extension of the present treatment to the case of double-barrier resonant tunneling structures may be quite worthwhile, in the sense that the trade-off between the transmission and the tunneling current valley polarization existing in one-barrier structures may be altered to suit applications.

## ACKNOWLEDGEMENTS

We would like to acknowledge the support of Ministry of Science and Technology in Taiwan through the Contract No. NSC103-2119-M-007-007 and the Thematic Project of Academia Sinica.

## APPENDIX

In **Sec. II**, it is stated for holes in an in-plane electric field that the major valley-dependent effects, namely, those due to TW and VOI, can be derived within the four-band $k \cdot p$ model given in Eq. (1). In this **Appendix**, we provide background details of the model as well as of the inclusion of SOC effects in the model. A similar and comprehensive theoretical treatment of the model with primary applications to conduction band electrons in a vertical electric field can be found in References 24 and 33.

For the construction of the $k \cdot p$ Hamiltonian, we use Bloch functions at high symmetry points (K and K′) as the basis set. In the spinless case, these Bloch functions are classified according to the irreducible representations $A_1$, $A_2$, $E_{1+}$, $E_{1-}$, $E_{2+}$, and $E_{2-}$ of the $C_{3h}$ group. For a reasonable $k \cdot p$ description of hole states near the valence band edge, we include a few bands around the valence band edge. Here, we start from a fourteen-band model [24] including the spin degeneracy and involving the following Bloch functions at K (K′) point: $|\psi^{c+2}_{E_{1-}(E_{1+})}, s\rangle$, $|\psi^{c+1}_{A_2(A_2)}, s\rangle$, $|\psi^{c}_{E_{1+}(E_{1-})}, s\rangle$, $|\psi^{v}_{A_1(A_1)}, s\rangle$, $|\psi^{v-1}_{E_{2-}(E_{2+})}, s\rangle$, $|\psi^{v-2}_{E_{2+}(E_{2-})}, s\rangle$, $|\psi^{v-3}_{E_{1-}(E_{1+})}, s\rangle$. Here, "s" denotes spin (s = +1/-1 or ↑/↓), the superscript denotes the band index, and the subscript denotes the corresponding irreducible representation. These states are chosen to cover as many distinct irreducible representations as possible so that the Hamiltonian is sufficiently general and manifests all qualitatively distinct inter-band couplings. The form of the $k \cdot p$ Hamiltonian can be determined using the symmetry of the basis functions. For example, under the $z \to -z$ reflection operation ($\sigma_h$), the basis functions are classified into the two following sectors, (i) the even sector given by $\{|\psi^{c+2}_{E_{1-}/E_{1+}}, s\rangle, |\psi^{c}_{E_{1+}/E_{1-}}, s\rangle, |\psi^{v}_{A_1/A_1}, s\rangle, |\psi^{v-3}_{E_{1-}/E_{1+}}, s\rangle\}$ and (ii) the odd sector given by $\{|\psi^{c+1}_{A_2/A_2}, s\rangle, |\psi^{v-1}_{E_{2-}/E_{2+}}, s\rangle, |\psi^{v-2}_{E_{2+}/E_{2-}}, s\rangle\}$. This leads to the fourteen-band Hamiltonian $H_{14b}$ with the following form:

$$H_{14b} = H_{kp} + H_{soc} + H_{ext}, \quad (A1)$$

where $H_{kp}$ is the Hamiltonian in the absence of SOC, $H_{soc}$ is the intrinsic SOC $\propto \nabla V \times \vec{p} \cdot \vec{S}$, and $H_{ext}$ is the potential energy due to an external electric field, given below:

$$H_{kp} = \begin{pmatrix} H_{even,\uparrow} & 0 & 0 & 0 \\ 0 & H_{even,\downarrow} & 0 & 0 \\ 0 & 0 & H_{odd,\uparrow} & 0 \\ 0 & 0 & 0 & H_{odd,\downarrow} \end{pmatrix} \quad (A1a)$$

$$H_{soc} = \begin{pmatrix} \Delta^{so}_{even,z} & 0 & 0 & \Delta^{so}_{-} \\ 0 & -\Delta^{so}_{even,z} & \Delta^{so}_{+} & 0 \\ 0 & (\Delta^{so}_{+})^{\dagger} & \Delta^{so}_{odd,z} & 0 \\ (\Delta^{so}_{-})^{\dagger} & 0 & 0 & -\Delta^{so}_{odd,z} \end{pmatrix} \quad (A1b)$$

$$H_{ext} = \begin{pmatrix} V_{ext} & 0 & \xi_z & 0 \\ 0 & V_{ext} & 0 & \xi_z \\ \xi_z^{\dagger} & 0 & V_{ext} & 0 \\ 0 & \xi_z^{\dagger} & 0 & V_{ext} \end{pmatrix} \quad (A1c)$$

$H_{kp}$ is block-diagonal, since the $k \cdot p$ term ($= k_x p_x + k_y p_y$) is even in the z-direction as well as spin-diagonal. $H_{even(odd),s}$ is the block Hamiltonian spanned by the basis functions in the even (odd) sector with spin s, with $H_{even(odd),\uparrow} = H_{even(odd),\downarrow}$. For $H_{soc}$, the terms proportional to $S_x$ and $S_y$ are odd in z, and mix the even and odd sectors with opposite spins giving $\Delta^{so}_{\pm} = \Delta^{so}_{x} \pm i\Delta^{so}_{y}$. In contrast, the terms proportional to $S_z$ only mix bands within the same sector and with the same spin giving $\Delta^{so}_{even,z}$ or $\Delta^{so}_{odd,z}$. For $H_{ext}$, the electron spin is conserved, and the out-of-plane field component couples bands between opposite sectors giving $\xi_z$ while $V_{ext}$ is taken to be slowly varying in the plane and thus contributes only to diagonal matrix elements [13].

Eq. (A1) shows that $H_{14b}$ is nearly diagonal. For the study of holes, it can be reduced to a smaller matrix involving less states by, for example, projecting $H_{14b}$ onto the even sector $\{|\psi^{c+2}_{E_{1-}/E_{1+}}, s\rangle, |\psi^{c}_{E_{1+}/E_{1-}}, s\rangle, |\psi^{v}_{A_1/A_1}, s\rangle, |\psi^{v-3}_{E_{1-}/E_{1+}}, s\rangle\}$ to which the valence band (v) belongs. In this projection, the off-diagonal couplings between v and those in the odd sector in $H_{14b}$ produces the second-order perturbation-theoretical terms that are proportional to $(|\Delta_\pm^{so}|)^2/\Delta_{BG}$, $|\Delta_\pm^{so}\xi_z|/\Delta_{BG}$, or $|\xi_z^2|/\Delta_{BG}$ ($\Delta_{BG}$ = typical band gap) and modify $H_{even,s}$. However, these terms are relatively small in comparison to those in $H_{even,s}$. For example, the term of $(|\Delta_\pm^{so}|)^2/\Delta_{BG}$ results in the spin at the valence band being tilted away from the z direction but, in comparison with $\Delta_{even,z}^{so}$ in $H_{even,s}$, it is lower in magnitude by $(\Delta_{so}/\Delta_{BG})^2$ ($\Delta_{so}$ = typical SOC matrix element in $H_{soc}$). Similarly, the term of $|\Delta_\pm^{so}\xi_z|/\Delta_{BG}$ can be shown to give rise to a Rashba SOC of the order of $kP_\mu^2\Delta_{so}(e\lambda_z)/\Delta_{BG}^3$ ($\lambda_z$ = the electric field in z-direction), which is lower than the VOI term in Eq. (2) by $\Delta_{so}/\Delta_{BG}$, if the in-plane and out-of-plane electric field components are taken to be of the same order. Therefore, for our study of the electric effect, it is ignored.

Accordingly, for the study of holes, we focus only on the even sector, or $H_{even,\uparrow}$ and $H_{even,\downarrow}$, and drop the rest blocks as well as the various couplings between the blocks. The forms of $H_{even,s}$ at K and K' valleys are different and related by the time-reversal operation. In terms of the valley index $\tau$ and spin s, they can be expressed as $H_{4b,s}^\tau = H_{even}^\tau + \tau s \Delta_{even,z}^{so}$ [17], summarized below:

$$H_{4b,s}^\tau = H_{4b} + H_{so},$$

$$H_{so} = \begin{pmatrix} \Delta_{so}^v & 0 & 0 & 0 \\ 0 & \Delta_{so}^{c+2} & 0 & \Delta_{so}^{c+2,v-3} \\ 0 & 0 & \Delta_{so}^c & 0 \\ 0 & \Delta_{so}^{c+2,v-3} & 0 & \Delta_{so}^{v-3} \end{pmatrix} \quad (A2)$$

$H_{4b}$ above is given by Eq. (1). For our study, the off-diagonal matrix elements in $H_{so}$ in Eq. (A2) can be ignored from the perturbation-theoretical viewpoint. Therefore, the effect of $H_{so}$ primarily shifts the various band edges by the amount of $\Delta_{so}$, i.e.,

$$E^v \to E^{v\prime} = E^v + \Delta_{so}^v, \quad E^{c+2} \to E^{c+2\prime} = E^{c+2} + \Delta_{so}^{c+2},$$
$$E^c \to E^{c\prime} = E^c + \Delta_{so}^c, \quad E^{v-3} \to E^{v-3\prime} = E^{v-3} + \Delta_{so}^{v-3}. \quad (A3)$$

With the Löwdin perturbation theory [35], we treat the off-diagonal $k \cdot p$ terms and diagonal $V_{ext}$ in (A2) as perturbations to the third order. This gives, for the topmost valence band, the following one-band, low-energy electron Hamiltonian

$$H_{1b}'(\tau) = H_0' + H_{TW}' + H_{VOI}'$$
$$H_0' = \frac{\hbar^2 k^2}{2(m^*)'} + V_{ext}$$
$$H_{TW}' = \tau\alpha_{TW}'(k_y^3 - 3k_yk_x^2)$$

$$H_{VOI}' = \tau\alpha_{VOI}'(\nabla V_{ext} \times \vec{k})\cdot\hat{z}, \quad (A4)$$

where $\alpha_{TW}'$, $\alpha_{VOI}'$, and $m^{*\prime}$ are basically those in Eqs. (2a)-(2c) except with the replacement prescribed in (A3). Therefore, the inclusion of SOC introduces a relative change of the order of $\Delta_{so}/\Delta_{BG}$ in the TW and VOI effects and the effective mass.